\documentstyle[aps,preprint]{revtex}

\title{Density matrix renormalization group for 19-vertex model}
\author{Yasushi Honda and Tsuyoshi Horiguchi \\
        Department of Computer and Mathematical Sciences, \\
        Graduate School of Information Sciences, 
        Tohoku University, Sendai 980-77, Japan}

\begin{document}
\maketitle

\begin{abstract}
  We embody the density matrix renormalization group method
for the 19-vertex model on a square lattice;
the 19-vertex model is regarded to be equivalent to the
XY model for small interaction.
The transfer matrix of the 19-vertex model is classified by
the total number of arrows incoming into one layer of the 
lattice. 
By using this property, 
we reduce the dimension of the transfer matrix 
appearing in the density matrix renormalizaion group method
and obtain a very nice value of the conformal anomaly
which are consistent with the value 
at the Berezinskii-Kosterlitz-Thouless transition point.


\end{abstract}
\newpage

The density matrix renormalization group method (DMRG) has been used
for evaluation of
eigenvalues of Hamiltonian matrix for a one dimensional quantum 
system \cite{White}.
This method enables us to increase the system size, 
within the matrix size 
which can be handled by recent computer resources such as the
memory size and the cpu speed. 
Recently the DMRG method has been
 applied to the transfer matrix of classical spin
system \cite{Nishino}.

In the both cases of
quantum Hamiltonian and classical transfer matrix,
spin variables have discrete degree of freedom.
Therefore one can make matrices with finite dimension in the 
both cases.
On the other hand,
a spin system with continuous degree of freedom such as
a classical XY model provides 
a transfer matrix whose dimension is infinite.
Fortunately it is known that the XY model 
on a square lattice $\Lambda$
is translated to a 19-vertex
model for which the transfer matrix can be described by discrete variables
expressed by arrow variables \cite{Knops}.
The purpose of the present study is to embody the DMRG method for the
19-vertex model.
We show how to reduce the dimension of matrices, which are
diagonalized in the
DMRG method, by using the ice rule of the 19-vertex model \cite{Baxtertext}.
We obtain very nice value of conformal anomaly, $c=1.006(1)$, at 
the Berezinskii-Kosterlitz-Thouless (BKT)
transition point by use of our method.

The 19 kinds of arrow configurations i.e. 
vertices are permitted by an ice 
rule which is generalized to include no arrow on a bond.
A vertex weight $W(v_i)$ depends on the kind of the vertex
$v_i \in \{1,2, \cdots, 19\}$ at site $i$.
The value of $v_i$ is determined by configuration of four arrows 
as follows:
\begin{equation}
  v_i = v_i(\alpha_i, \beta_i, \gamma_i, \delta_i) ,
  \label{eq:vertex}
\end{equation}

where $\alpha_i, \beta_i, \gamma_i$ and $\delta_i$ denote arrows
on bonds surrounding the site $i$.
Let us express an up and a right arrow by $+1$, 
a down and a left arrow by $-1$,
no arrow by $0$.
For instance, we have $v_i(-1,0,0,+1)=1$ as shown in Fig.1

of ref.\cite{Knops}.
The ice rule is expressed by using this
expression as follows:
\begin{equation}
  \alpha_i-\beta_i-\gamma_i+\delta_i = 0 .
\label{eq:ice}
\end{equation}
Using the weights $W(v_i)$ for 19 vertices, we can describe the 
partition
function $Z$ as follows:
\begin{equation}
  Z={\sum_{\{ v_i \}}}' \prod_{i \in \Lambda} W(v_i) ,
\label{eq:19vertexptf}
\end{equation}
where the summation is taken over all permitted configurations
of the vertices on the lattice $\Lambda$.

Because there exists the ice rule for the 19-vertex model, the 
whole transfer matrix becomes a block diagonal which is 
classified by the number of arrows incoming to one layer of the lattice.
By using this property, we can reduce the amount of calculation
in the DMRG method.

We now apply the infinite DMRG method to the 19-vertex
model \cite{White}.
In addition to the vertex weight $W(v_i)$, 
we use a renormalized weight $W^{(r)}(v_i^{(r)})$,
where $v_i^{(r)}$ means a renormalized vertex and $r$ the number
of a renormalization.
As an initial value, $W^{(0)}(v_i^{(0)})$ is equal to $W(v_i)$.
The transfer matrix for $N$, the number of arrows, is composed as
\begin{eqnarray}
  & & \hspace{-10mm} 
  T_N^{(r)}(\eta_1,\beta_2,\eta_3,\beta_4|\xi_1,\delta_2,\xi_3,\delta_4) 
\hspace*{-9mm}
\nonumber \\
  &=& \hspace{-4mm} \sum_{\alpha_1,\cdots,\alpha4} \hspace{-3mm}
          W^{(r)}(v_1^{(r)}) W(v_2)  W^{(r)}(v_3^{(r)}) W(v_4)  ,
 \label{eq:maket}
\end{eqnarray} 
where the number $N$ is obtained by
\begin{eqnarray}
  N &=& N_r(\eta_1)+\beta_2+N_r(\eta_3)+\beta_4 \nonumber \\
    &=& N_r(\xi_1)+\delta_2+N_r(\xi_3)+\delta_4 .
  \label{eq:Nconsv}
\end{eqnarray}
Here the number of arrows included in the renormalized vertex $\xi_i$ 
is denoted by $N_r(\xi_i)$ and is equal to $\delta_i$ in the 
initial step of the DMRG procedure. 
We denote an eigenvector of this transfer matrix by
$\psi_{N,k}^{(r)}(\eta_1,\beta_2,\eta_3,\beta_4)$ which corresponds to 
the $k$-th eigenvalue.
As the infinite DMRG method, we construct the density matrix
$\hat{\rho}_{N_r(\eta_1)+\beta_2,k}^{(r)}$
as follows:
\begin{eqnarray}
   & & \hspace{-9mm}
   \rho_{N_r(\eta_1)+\beta_2,k}^{(r)}(\eta_1,\beta_2|\xi_1,\delta_2) 
   \nonumber \\
     \hspace{-3mm} &\equiv& \hspace{-3mm} \sum_{\eta_3,\beta_4} 
     \psi_{N,k}^{(r)}(\eta_1,\beta_2,\eta_3,\beta_4) 
     \psi_{N,k}^{(r)}(\xi_1,\delta_2,\eta_3,\beta_4)  .
\end{eqnarray}
Notice that the density matrix is labeled by $N_r(\eta_1)+\beta_2$
not by $N$.
From Eq.(\ref{eq:Nconsv}), 
since $\eta_3=\xi_3$ and $\beta_4=\delta_4$,
$N_r(\eta_1)+\beta_2 = N_r(\xi_1)+\delta_2$,
therefore the density matrix has a block diagonal structure classified by
the total number of arrows for the half system.
In order to construct the renormalized vertex weight $W^{(r+1)}$,
we diagonalize the density matrix $\hat{\rho}_{N_r(\eta_1)+\beta_2,1}$
and obtain its eigenvectors 
$\vec{V}_{N_r(\eta_1)+\beta_2,\eta'_1}$.
In the present study, we use the eigenvector of the transfer 
matrix for the largest eigenvalue $k=1$.
The renormalized vertex state is labeled by $\eta'_1$ which means
that $\vec{V}_{N_r(\eta_1)+\beta_2,\eta'_1}$ corresponds to the 
$\eta'_1$-th eigenvalue of $\hat{\rho}_{N_r(\eta_1)+\beta_2,1}$.

We determine the upper limit $l$ of $\eta'_1$ as follows:
\begin{eqnarray}
  l \equiv \left\{
     \begin{array}{ll} 
        3^{r+2} & (3^{r+2}<m) \\
        m       & (3^{r+2}\geq m)  ,
     \end{array}
  \right.
  \label{eq:rglimit}
\end{eqnarray}
where $m$ is the number of states of the density matrix which we take
into account.
The last step of the DMRG for the 19-vertex model is a construction 
of the renormalized weight for the vertex as follows:
\begin{eqnarray}
  & & \hspace*{-2em} 
   W^{(r+1)}(v_1^{(r+1)}(\alpha_1,\eta'_1,\alpha_3,\xi'_1)) 
   \nonumber \\
  &=& \hspace{-1em}
      \sum_{\alpha_2,\eta_1,\beta_2,\xi_1,\delta_2} \hspace{-1em}
      V_{N_r(\eta'_1),\eta'_1}(\eta_1,\beta_2) 
      W^{(r)}(v_1^{(r)}(\alpha_1,\eta_1,\alpha_2,\xi_1))
      \nonumber \\
  & & \times  W(v_2(\alpha_2,\beta_2,\alpha_3,\delta_2)) 
              V_{N_r(\xi'_1),\xi'_1}(\xi_1,\delta_2) .
\end{eqnarray}
The total number of arrows included in the renormarized vertex becomes
\begin{equation}
  N_r(\eta'_1) = N_r(\xi'_1) = N_r(\xi_1)+\delta_2  .
\end{equation}
We need $W^{(r+1)}(v_3^{(r+1)}(\alpha_3,\eta'_3,\alpha_4,\xi'_3))$
to return to the first step of the DMRG method, but we do not need
to calculate it.
Since in our method for the 19-vertex model 
the system has a symmetry of 
translation, we can use $W^{(r+1)}(v^{(r+1)}_1)$ as $W^{(r+1)}(v^{(r+1)}_3)$.
Then we return to the first step represented by Eq.(\ref{eq:maket})
to iterate the DMRG procedure.
By iterating the above DMRG procedure for the 19-vertex model with
the restriction of the total number
of arrows as mentioned here,
we can make the system size $L$ increase systematically.
The advantage of the present
method is that the dimension of the transfer matrix
decreases dramatically by considering the conservation law of the number
of arrows in this DMRG method.
For example in the case of $N=0, m=35$ and $L=12$,
the dimension of matrix is dramatically reduced from 531441 down to 1545.

The value of the conformal anomaly has to be 1 for the
BKT transition point.
The result at $K=1.0866$ is shown in Fig.1.
The value of interaction is equal to 
that estimated as the critical value in
ref.\cite{Knops}.
We obtain $c=1.006(1)$ which is consistent with the value
for the BKT point.

As summary, we have succeeded to reduce the matrix dimension to be
diagonalized in the DMRG method for the 19-vertex model by
using the ice rule.
A very nice value of the conformal anomaly is obtained  
at BKT transition point by the present approach.

\vspace*{10mm}
Figure caption \\
{\label{fig:cinv}Fig.1 The size dependence of the free energy.}

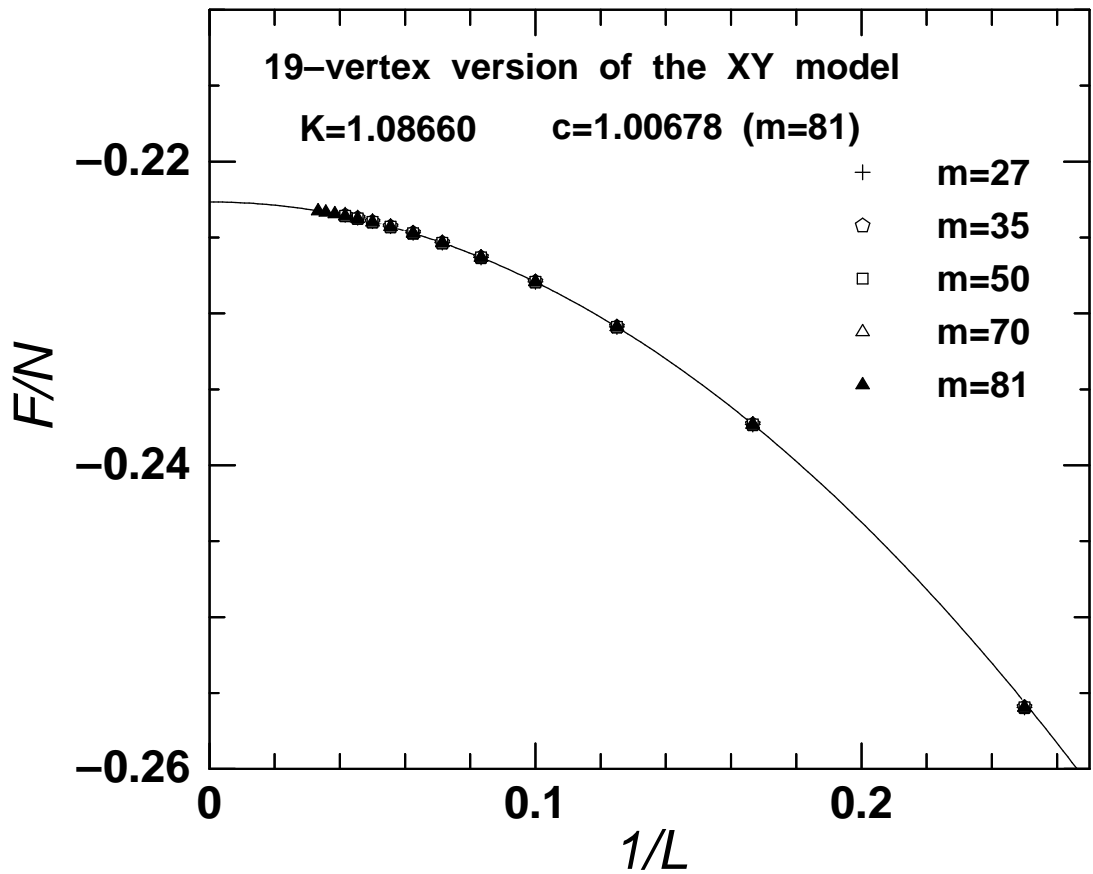
\end{document}